\documentclass[journal=jpcafh,manuscript=article,layout=twocolumn]{achemso}
\usepackage{amsmath}

\title[Positive Charge in Superfluid $^4$He]{Solvation of Intrinsic Positive Charge in Superfluid Helium}

\author{David Mateo}
\affiliation{Department of Chemistry and Biochemistry, California State University at Northridge, 18111 Nordhoff St., Northridge, CA 91330, USA}

\author{Jussi Eloranta}
\email{Jussi.Eloranta@csun.edu}
\affiliation{Department of Chemistry and Biochemistry, California State University at Northridge, 18111 Nordhoff St., Northridge, CA 91330, USA}


\begin{document}

\twocolumn[
\begin{@twocolumnfalse}
\begin{abstract}
Based on electronic structure calculations, the structure of intrinsic positive charge solvated in superfluid helium is identified as triatomic He$_3^+$ ion, which is bound to the surrounding ground state helium atoms through the charge - charge induced dipole interaction in a pairwise additive manner. Bosonic density functional theory calculations show that this ion forms the well-known Atkins' snowball solvation structure where the first rigid helium shell is effectively disconnected from the rest of the liquid. Evaluation of the total energy vs. helium droplet size $N$ shows distinct regions related to the completion of solvent shells near $N=16$ and $N=47$. These regions can be assigned to magic numbers observed in positively charged helium droplets appearing at $N=15$ and in the range between $20-50$ helium atoms. The calculated added mass for the positive ion in bulk superfluid helium (18 $m_{He}$) is much smaller than the previous experiments suggest (30 -- 40 $m_{He}$) indicating that there may be yet some unidentified additional factor contributing to the measured effective mass. Both previous experiments and the present calculations agree on the effective mass of the negative charge (240 -- 250 $m_{He}$). The main difference between the solvated negative and positive charges in liquid helium is that the latter forms a chemically bound triatomic molecule surrounded by highly inhomogeneous liquid structure whereas the former remains as a separated charge with a smoothly varying liquid density around it.
\end{abstract}
\end{@twocolumnfalse}
]

\section{Introduction}

Positively charged helium clusters, He$_n^+$, have been extensively studied in the gas phase and helium droplets \cite{northby1,grandinetti1}. The early gas phase experiments focused on He$^+$ and He$_2^+$ but, subsequently, Patterson was able to identify He$_3^+$ for the first time in his ion mobility experiments below 200 K \cite{patterson1}. By analyzing the temperature dependency of the equilibrium constant for He$_2^+$ + He $\rightleftharpoons$ He$_3^+$ reaction, he was able to determine the dissociation enthalpy, $\Delta_r H^\circ$, for He$_3^+$ as $-16.4\pm 3$ kJ/mol.\cite{patterson1} Later Kobayashi \textit{et al.} showed that the binding energy of neutral He atoms to He$_n^+$ clusters decreases significantly after $n = 3$.\cite{kaneko1} This indicates that the larger clusters consist of a He$_3^+$ core with the remaining neutral He atoms bound to it by the charge - induced dipole interaction and repelled at short distances by the Pauli exclusion interaction. This picture was further confirmed by the experiments of Hirioka \textit{et al.} where the reaction enthalpies and entropies for clustering reactions, He$^+_n$ + He $\rightleftharpoons$ He$^+_{n+1}$, were studied systematically.\cite{mori1} The measured reaction enthalpies show a sudden decrease after the formation of the triatomic ion core is complete (i.e., $\Delta_r H^\circ = -15.6$ kJ/mol for $n=2$ and $\Delta_r H^\circ \approx -2.7$ kJ/mol for $n>2$) whereas the corresponding reaction entropies have much weaker dependency on $n$.\cite{mori1} Spectroscopic studies of Haberland \textit{et al.} identified the He$_3^+$ absorption band $X$ $^2\Sigma_g^+ \rightarrow 1$ $^2\Sigma_u^+$ near 5.3 eV, which was observed to red shift slightly as the cluster size increased.\cite{toennies1} This observation implies that the valence electronic structure of the triatomic ion was also preserved in lager helium clusters.\cite{toennies1} Interestingly, the metastable quartet state of He$_4^+$ was also seen in these experiments.\cite{toennies2,murrell1} 

Numerous experimental studies exist in the literature where the solvated positive charge (He$_n^+$) was used as a probe to study ion mobility in superfluid helium (i.e., the viscous response due to phonon/roton/$^3$He scattering)\cite{reif1,reif2,reif3,meyer1,borghesani1}, roton/vortex nucleation when the critical velocity is exceeded \cite{meyer1,reif4,santini1,borghesani1}, and the interaction of impurities with quantized vortex lines \cite{reif4,rayfield1,borghesani1}. Most of these experiments were designed to study the atomic scale response of bulk superfluid helium rather than to characterize the solvated positive ion itself. However, two important experimental techniques have been applied to determine the effective mass of the positive ion in bulk superfluid helium: 1) the microwave loss technique of Dahm and Sanders\cite{sanders1} and 2) the microwave resonance technique of Poitrenaud and Williams\cite{williams1,williams2}. Both methods predict that the positive ion effective mass corresponds to 30 -- 40 helium atoms whereas the negative charge (i.e., solvated electron) is much heavier, \textit{ca.} $240 m_{He}$.\cite{sanders1,williams2} A theoretical estimate based on the Atkins' snowball model indicates that the effective mass of the positive ion in the bulk consists of 40 -- 60 He atoms that are rigidly attached to the ion giving it an effective radius of 6~\AA{} \cite{atkins1,borghesani1}. In this model, the interaction between the positive ion and the surrounding helium atoms is approximated only by the charge induced polarization interaction, which implies that all positive ions (e.g., Ba$^+$, Ca$^+$) would interact with helium the same way. However, considering the differences in the electronic structure of the various positive ions, it becomes clear that this must be a rather crude approximation. This was clearly demonstrated by experimental ion mobility studies where the mobilities were shown to depend strongly on the particular positive ion used\cite{johnson1} and by quantum Monte Carlo calculations of alkali metal ions solvated in superfluid helium that predicted the formation of a complex solvation structure with varying degree of helium localization around the ions.\cite{reatto1} Since the expected core structure of the positive ion in superfluid helium is He$_3^+$, as discussed above, it is clear that the simple charge induced polarization model is not sufficient to describe the ion - helium interaction accurately. Furthermore, the Atkins' model does not provide any specifics about the strongly inhomogeneous liquid structure near the ion, which is expected to result in a complex solvation shell structure extending several \AA{}ngstr\"oms away from the ion, nor considers the hydrodynamic liquid response to the complex. Refined models have been developed to account for the relative mobilities of different species,\cite{bachman1} but they do not include the microscopic details of ionic solvation explicitly.

Due to the small number of electrons present in He$_n^+$ ions ($n<10$), it has been possible to use high level \textit{ab initio} electronic structure calculations to model their energetics and geometries with high accuracy. One of the first successful calculations for He$_n^+$ ($n = 2-7$) was reported by Rosi \textit{et al.} where they used the Hartree-Fock (HF) method along with the modified coupled pair functional (MCPF) theory and a specially tailored basis set combined with the atomic natural orbital (ANO) approach \cite{bauschlicher1}. These calculations were able to provide approximately the correct equilibrium geometries but did not provide very accurate energetics. Especially, it should be noted that the stated agreement by Hiraoka and Mori between the experimental results\cite{mori1} and the MCPF/ANO calculations\cite{bauschlicher1} was just coincidental because the experimental reaction enthalpies and the calculated changes in electronic energy for the ion formation cannot be directly compared. If the zero-point corrections were included from the HF results of Rosi \textit{et al.},\cite{bauschlicher1} the MCPF/ANO calculation would only account for approximately half of the experimentally obtained binding energies for He$^+_n$ + He $\rightleftharpoons$ He$^+_{n+1}$ when $n>2$.\cite{mori1} Furthermore, the thermal contributions of translational, rotational and vibrational degrees of freedom to the reaction enthalpies were neglected. The more recent electronic structure calculations for small ground state He$_n^+$ clusters have employed variants of the configuration interaction method (CI) or the coupled clusters theory (CC) with the latter including single (S), double (D) and perturbative triples excitations (T), and typically used a medium sized correlation consistent basis set \cite{gianturco1,gellene1,gianturco2,yurtsever1}. The main aim of these studies has been to characterize the full potential energy surface for ground state He$_3^+$ and to compute the bound He$_3^+$ vibronic states with respect to the He$_2^+$ + He dissociation channel \cite{gianturco1,gellene1,gianturco2}. From the perspective of the triatomic ion formation, the barrierless linear He$_2^+$ -- He approach leads directly to the formation of He$_3^+$. When the zero-point energies were included in the energetics, these calculations were able to reproduce the experimentally observed dissociation energies for He$^+_2$ + He $\rightleftharpoons$ He$^+_3$ within 1 kJ/mol\cite{mori1,gianturco2}. The calculation involved solving the full triatomic vibrational problem numerically and therefore the obtained vibronic eigenstates (and the zero-point correction) were not subject to the harmonic approximation \cite{gianturco2,gellene2}. Despite of the obtained good agreement with the experimental data at 110 K, the thermal contributions to $\Delta_r H^\circ$ were not included.\cite{mori1} The recent work of Marinetti \textit{et al.} have considered larger He$_n^+$ ($n>3$) by using the M\o{}ller-Plesset MP4(SDQ) level of theory with a triple zeta level correlation consistent basis set to obtain the equilibrium structures and energies \cite{yurtsever1}. The results from these calculations are consistent with the experimental observation that all He$_n^+$ ($n>3$) have a triatomic ionic core, which is surrounded by ground state He atoms bound to it by electrostatic interaction. The potential energy surface for He$_3^+$ -- He appears deeply bound near the ion with the T-shape geometry having the lowest energy ($-3.5$ kJ/mol). Most importantly, it was observed that the addition of more He atoms around the ion followed approximately a pair-wise additive behavior such that the total interaction could be obtained simply by summing over the ion core -- He pairs as well as the He -- He pairs that do not belong to the ionic core \cite{yurtsever1}. Finally, we note the series of diatomics-in-molecules (DIM) studies of charged helium clusters, where the basic DIM method surprisingly breaks down for the relative simple He$_3^+$ molecule \cite{murrell1,apkarian1,kalus1,gadea1}. DIM is an attractive alternative for  the computationally heavy traditional \textit{ab initio} methods and is, for example, fast enough to be used in molecular dynamics work \cite{eloranta1}.

In the first part of this paper, we apply the CCSD(T)/AV6Z level electronic structure theory to obtain the equilibrium structures for He$_2^+$, He$^+_3$, He$^+_4$ and He$^+_5$, compare the calculated energetics and vibrational frequencies with the literature values, and generate the global potential energy surface for He$_3^+$ -- He. In the second part, we employ the Orsay-Trento bosonic density functional theory (DFT) to describe the solvation of He$_3^+$ as a function of increasing helium droplet size, finally converging towards the bulk liquid limit. The equilibrium liquid solvation structures were obtained and the rotational behavior of He$_3^+$ in the bulk liquid was classified (i.e., stopped vs. free rotor). Finally, we estimate the effective hydrodynamic mass of He$_3^+$ in bulk superfluid helium and discuss its relationship to previous experimental and theoretical results. 

\section{The Computational Approach}

The electronic structure calculations of He$_n^+$ were carried out at the spin restricted CCSD(T) level of theory with an augmented correlation consistent basis set, aug-CC-pV6Z (AV6Z) \cite{molpro1,peterson2}. The equilibrium geometries were obtained by the standard geometry optimization module as implemented in the Molpro code \cite{molpro2}, followed by the calculation of harmonic vibrational frequencies. The standard $T_1$-norm test was carried out in each calculation to ensure that the contribution of single excitations to the reference wavefunction was not excessive \cite{taylor1}. Even though AV6Z is very close to a complete basis set, the potential energy surface calculations for He$_3^+$ - He included the basis set superposition error (BSSE) correction through the counterpoise method of Boys and Bernardi \cite{bernardi1}. The maximum difference between the BSSE corrected and uncorrected binding energies was typically less than 0.08 meV. The complete potential energy surface for He$_3^+$(linear molecule) -- He was generated from five different angular cuts with angles 0$^\circ$ (linear geometry), 22.5$^\circ$, 45$^\circ$, 67.5$^\circ$ and 90$^\circ$ (T geometry) by using interpolation for the full potential energy surface (see Appendix). The thermodynamic functions, enthalpy and entropy, were evaluated at previously optimized geometries by taking harmonic behavior for the vibrational motion, assuming ideal gas behavior for the translational degrees of freedom and removing the symmetry forbidden rotational states ($^4$He are bosons) from the calculation of the rigid rotor partition function \cite{landau1}. The standard reaction enthalpies ($\Delta_r H^\circ$) and entropies ($\Delta_r S^\circ$) are reported for reactions He$^+_n$ + He $\rightleftharpoons$ He$^+_{n+1}$ within the assumptions stated above. The numerically exact fundamental vibrational frequencies for He$_3^+$ were obtained by the vibrational self-consistent field (VSCF) method followed by a vibrational configuration interaction (VCI/SDTQ) calculation \cite{rauhut1,rauhut2,rauhut3,rauhut4,gerber1}. 

The liquid helium surrounding He$_3^+$ was modeled by the Orsay-Trento DFT (OT-DFT) method where the He$^+_3$ -- He pair potential obtained above acted as an external potential for the liquid \cite{treiner1}. Since the atomic binding energies in the He$_3^+$ core are several orders of magnitude larger than the ground state He - He attractive interaction, the treatment of He$_3^+$ as a separate entity from the surrounding liquid is justified. Within this model, the solvation energy of He$^+_3$ and the equilibrium liquid distribution around it can be found by minimizing the total free-energy of the system, i.e. solving
\begin{equation}
\frac{\delta}{\delta\Psi^*_{liq}(r)}\bigg(E[\Psi_{liq}] - \mu\int|\Psi_{liq}|^2 \bigg) = 0
\qquad ,
\label{min_eq}
\end{equation}
where
\begin{align}
&E[\Psi_{liq}] = \nonumber \\
&\int \left\{
        \frac{\hbar^2}{2m}|\nabla\Psi_{liq}|^2
        + \epsilon_{OT}[\Psi_{liq}]
        + V_{He^+_3-He}|\Psi_{liq}|^2 
\right\}
\label{total_e}
\end{align}
and $\Psi_{liq}$ is the liquid helium order parameter (``effective wavefunction''), $m_{He}$ is the helium atom mass, $\epsilon_{OT}$ represents the OT energy density functional,\cite{treiner1} $V_{He^+_3-He}$ is the computed (anisotropic) He$^+_3$ - He pair potential and $\mu=-7.15$ K/atom is the chemical potential at the saturated vapor pressure and zero temperature. Due to the strongly bound nature of the external potential, an additional term to account for possible solidification of helium was included in OT-DFT \cite{pi1}. This modified functional is known to produce spontaneous symmetry breaking in some situations where localized regions of high density appear around an attractive impurity forming a solid-like structure. When this approach was used to study alkali ions, we observed that taking a spherical average of the liquid density led to consistent results with quantum Monte Carlo calculations.\cite{toigo1,eloranta2}

We have solved the non-linear Schr\"odinger-type equation corresponding to Eq.~(\ref{min_eq}) by means of imaginary time propagation.\cite{eloranta3} The numerical treatment of the helium OT-DFT problem is described elsewhere.\cite{eloranta4} To minimize the boundary condition artifacts arising from using a finite sized simulation box, a large grid consisting of $128\times128\times128$ points with a spatial grid step of $0.25$ \AA{} was used in the calculations. An imaginary time step of 1 fs was used in minimizing the total energy of the system. To verify that there was no time step bias in the obtained solution, shorter time steps down to 0.01 fs were executed at the end of each run. Typically $10^4$ imaginary time iterations were required for full convergence.

Due to its interaction with the liquid, the ion travels through the medium with an effective mass $m^* = m_{ion} + m_{add}$. The hydrodynamic added mass $m_{add}$ can be calculated from the second derivative of the energy with respect to the ion velocity.\cite{lehmann1} To obtain the minimum quasienergy configuration where the ion moves at a constant velocity $\vec{v}_0$, we solve Eq.~(\ref{min_eq}) in the frame of reference co-moving with the ion. In practice, this is done by including an additional term to the energy functional such that\cite{pi2}
\begin{equation}
E[\Psi_{liq}, \vec{v}_0] =
E[\Psi_{liq}, 0] - \vec{v}_0 \cdot \int \Psi_{liq}^*(r) \vec{P} \Psi_{liq}(r)d^3r
\label{frame_e}
\end{equation}
where $\vec{P}$ represents the momentum operator for $^4$He. Alternatively, we also evaluated the added mass by computing, in the co-moving frame,
\begin{align}
\frac{m_{add}}{m_{He}} &= \frac{\vec{v}_0}{v^2_{0}}\int \rho(r)\left(\vec{v}(r) + \vec{v}_{0} \right)d^3r \nonumber \\
&= N + \frac{1}{v^2_{0}} \int \vec{v}_0\cdot\vec{\jmath}(r) d^3r
\label{momentum}
\end{align}
where $\vec{v}$ is the liquid velocity, $\vec{\jmath}$ is the density current and $N$ is the number of helium atoms in the simulation box. Note that in the co-moving frame $\vec{v}_0\cdot\vec{\jmath}\le0$. Both methods yielded essentially identical results as real-time propagation of the system under a constant electric field presented in our previous work\cite{eloranta2} but are computationally less demanding.
The added mass can be approximately related to the hydrodynamic radius $R_b$ of the ion by \cite{eloranta2}:
\begin{equation}
m_{add} = m_{He}\int_0^{R_b}\tilde{\rho}(r)d^3r
\label{eq3}
\end{equation}
\begin{equation}
\tilde{\rho}(r) = \left\lbrace
\begin{array}{cl}
(\rho_0 + \rho(r))/2	& \textnormal{ when }\rho(r) < \rho_0\\
\rho(r) 				& \textnormal{ when }\rho(r) \ge \rho_0
\end{array}
\right.
\label{eq4}
\end{equation}

\noindent
where $\rho(r)$ represents the liquid density at point $r$ and $\rho_0$ is the bulk liquid density.

\section{Results and Discussion}

\subsection{Electronic structure calculations}

The calculated equilibrium geometry data and harmonic vibrational frequencies for He$_n^+$ clusters up to $n=5$ are collected in Table \ref{table1}. The obtained He$_3^+$ core bond length, $R_1$, is essentially identical to what was reported previously using MP4(SDQ)/AVTZ and QCISD(T)/AVQZ levels of theory.\cite{gianturco1,yurtsever1,gellene2} Slightly larger deviations from the earlier calculations are seen along the weakly bound coordinate, $R_2$, that is oriented along the He$_3^+$ -- He direction. Binding along this coordinate is mostly due to the charge (He$_3^+$) -- charge induced dipole (polarized He atom) interaction, which is expected to be at maximum in the T shaped geometry ($\theta = 90^\circ$). The general trend is that the present calculations produce slightly shorter bond lengths, which can be attributed to the larger basis set used in this study (AV6Z) vs. the smaller correlation consistent basis sets in earlier studies (AVTZ and AVQZ)\cite{gianturco1,yurtsever1,herzberg1,gellene2}. Based on the calculations, the linear He$_2^+$ -- He coordinate is barrierless and such approach may lead to the formation of He$_3^+$ directly. This channel is expected to be very efficient at low temperatures and high helium densities, which is consistent with the experimental conditions where He$_3^+$ has been observed previously. The calculated potential energy surfaces for clusters with $n>3$ display anharmonic behavior along the $R_2$ coordinate, which can lead to large errors in vibrational frequencies when obtained by using the harmonic approximation. Thus, at most the calculated harmonic frequencies in this case can only be used as a consistency check of the local potential energy surface behavior near the minimum energy geometry as compared with previous theoretical studies. As shown in Table \ref{table1}, the calculated He$_2^+$ and He$_3^+$ harmonic frequencies agree with the previous literature data very well \cite{herzberg1,gellene2}. In order to go beyond the harmonic approximation, we have also used the VSCF/VCI method at CCSD(T)/AV6Z level of theory to compute the fundamental frequencies for He$_3^+$ within $D_{2h}$ point group. These frequencies are more sensitive to the non-local behavior of the potential energy surface near the energy minimum and can therefore offer a better indicator of the behavior of the calculated surface in this region. Our calculations predict the frequencies as follows: 715 cm$^{-1}$ ($A_g$; symmetric stretch), 452 cm$^{-1}$ ($B_{1u}$; asymmetric stretch) and 230 cm$^{-1}$ ($B_{2u}/B_{3u}$; bending). These values are fairly close to those obtained by Satterwhite \textit{et al.}\cite{gellene2}: 643 cm$^{-1}$ ($A_g$), 441 cm$^{-1}$ ($B_{1u}$) and 234 cm$^{-1}$ ($B_{2u}/B_{3u}$) but appear to deviate more from the results of Scifoni and Gianturco\cite{gianturco2}: 564 cm$^{-1}$ ($A_g$), 382 cm$^{-1}$ ($B_{1u}$) and 450 cm$^{-1}$ ($B_{2u}/B_{3u}$). It is unclear, if in the latter case the deviation is due to differences present in the potential energy surface or solving the triatomic vibrational problem numerically. The VSCF/VCI calculation gives a total zero-point energy of 950 cm$^{-1}$, which is very close to the harmonic value of 946 cm$^{-1}$. We conclude that the present electronic structure calculations produce consistent results with the existing literature data for small He$_n^+$ clusters and the use of the large AV6Z basis set led only to a minor improvement of the potential energy surface over the previous calculations.\cite{gianturco1,yurtsever1,gellene2}

For the present consideration, the two most important outcomes from the above calculations are: 1) the formation of He$_3^+$ is spontaneous along the linear approach and 2) all He$_n^+$ clusters with $n>3$ preserve the triatomic He$_3^+$ core to which the rest of the ground state He atoms are bound to. The latter statement can be verified by inspecting the data shown in Table \ref{table2} where the calculated helium atom binding energies of He$_n^+$ clusters are reported and compared with the existing literature data \cite{mori1}. We have also calculated the standard thermodynamic formation enthalpies ($\Delta_r H^\circ$) and entropies ($\Delta_r S^\circ$) for He$_n^+$ + He $\rightleftharpoons$ He$_{n+1}^+$ equilibrium reaction as shown in Table \ref{table2}. By comparing the calculated values against the experimental data\cite{mori1}, it is apparent that our calculations overestimate the formation enthalpies by \textit{ca.} 1 - 2 kJ/mol. This may be related in part to the harmonic approximation used, neglecting the possible coupling between vibrational modes and ignoring the coupling between the vibronic and rotational degrees of freedom, which all can contribute to $\Delta_r H^\circ$ and $\Delta_r S^\circ$. However, at least the first source of error can be safely discarded since the total vibrational partition function is very small at low temperatures, $T<110$ K. We eliminated the symmetry forbidden rotational states (i.e., for $^4$He only even rotational states are allowed for $g$ symmetric electronic states and odd for $u$ symmetric) from the evaluation of the relevant partition functions, which led to, for example, \textit{ca.} 5 J / (mol K) increase in the reaction entropies ($n = 2,3$). The correction appeared to work very well for $n = 3$ (see Table \ref{table2}) where the calculated and experimental values agree within the experimental error bar. Unfortunately,  it is very difficult to carry out such correction exactly for $n=4$ since it involves an asymmetric top rotor (i.e., all the three moments of inertia are different)\cite{herzberg1} and hence we only give a lower limit for $\Delta_r S^\circ$ without eliminating the contribution from the forbidden rotational states. Overall, a much better agreement with the experimental formation enthalpies is achieved if only the zero-point vibrational corrections are included (i.e., taking the $T=0$ K limit). These values show agreement with experiments better than 0.4 kJ/mol. For the zero-point corrected electronic energies, note in particular the large increase in the He binding energy when going from He$^+_2$ to He$_3^+$ and that for $n\ge 3$ the increase in binding energy becomes independent of $n$. As discussed earlier, the former observation means that the persistent form of intrinsic positive charge in any low temperature dense helium system is He$_3^+$ whereas the latter suggests that one can construct the potential energy surface for the He$_3^+$ -- He$_n$ system by using a pair-wise additive potential model. The pair-wise additivity was analyzed and discussed in more detail by Marinetti \textit{et al.}\cite{yurtsever1} and our results support their findings for $n>3$. 

\begin{figure}
\includegraphics[width=\linewidth]{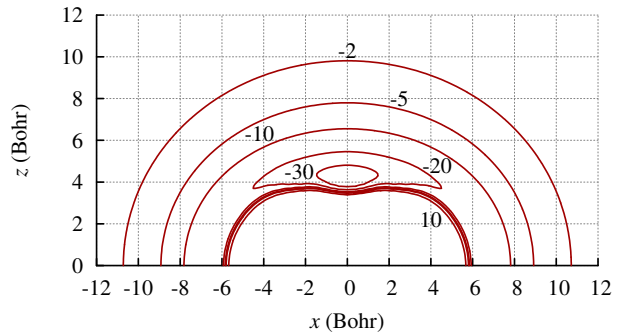}
\caption{He$_3^+$--He isopotential lines (meV) for He$^+_3$ molecule oriented along the $x$ axis. The units were chosen to facilitate comparison with the results given in Ref.~\citenum{yurtsever1}.}\label{he3pot}
\end{figure}

Based on the above discussion, solvation of the intrinsic positive charge in liquid helium (He$_n^+$) can be modeled by using a pair-wise additive model, which is based on the calculated He$_3^+$ - He potential energy surface and the van der Waals interaction between the surrounding ground state He atoms. This form of potential energy function can be directly applied in theoretical treatments that can account for the quantum mechanical nature of the weakly bound He atoms surrounding the ionic core (e.g., quantum Monte Carlo, OT-DFT). The full He$_3^+$ -- He potential energy surface (raw data given in the Appendix) was generated by interpolating along both the distance $R_2$ from the center of mass of He$_3^+$ and the angle $\theta$ between the linear molecule and the ground state He atom. As shown in Fig. \ref{he3pot}, this surface is nearly identical to that obtained earlier by Marinetti \textit{et al.}\cite{yurtsever1} It exhibits largest attraction in the T shape geometry ($\theta = 90^\circ$) at \textit{ca.} $R_2 = 2.4$ \AA{} with a binding energy of 438 K. This geometry can be rationalized by noting that the T approach has a better access to the positive core than the linear geometry (screening of charge) and hence the charge - charge induced dipole interaction is maximized. Overall the bound region of the potential extends over all angles and, for example, in the linear geometry ($\theta = 0^\circ$) it still reaches 232 K at 3.4 \AA{} distance. Due to this strong short-range electrostatic interaction, we expect that strongly bound solvation shells form around the ion in a dense helium environment.

\subsection{He$_3^+$ rotational degrees of freedom}

\begin{figure}
\includegraphics[width=\linewidth]{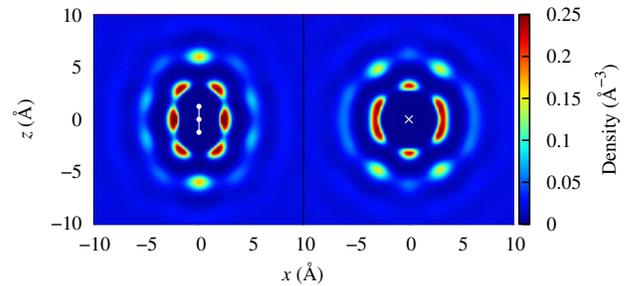}
\caption{Calculated helium density around a solvated He$^+_3$ ion: a non-rotating ion (left) and a freely rotating ion (right).}\label{free-fix}
\end{figure}

When He$_3^+$ is directly introduced in OT-DFT as an external potential, the zero-point motion for both the translational and rotational degrees of freedom for the ion are neglected. In a set of test calculations, the translational zero-point motion was included\cite{eloranta6} and it was observed that this effect can be safely ignored for the present system. For molecular rotation, we consider the energetics of two extreme cases: 1) freely rotating molecule ($J=0$) where the rotation dominates over the interaction with the liquid and 2) stopped rotor (i.e., non-rotating linear molecule) where the anisotropic molecule - liquid potential freezes the rotational motion. A freely rotating molecule in a $J=0$ state can be modeled by replacing the anisotropic pair-potential in Eq.~(\ref{total_e}) by its spherical average
\begin{align}
&\bar{V}_{He^+_3-He}(|\vec{r}-\vec{r}_{ion}|) = \nonumber \\
&\frac{1}{2}\int_0^\pi \sin\theta_{ion} V_{He^+_3-He}(\vec{r}-\vec{r}_{ion},\theta_{ion})d\theta_{ion}
\quad ,
\end{align}
where $\vec{r}_{ion}$ is the position of the ion and $\theta_{ion}$ its angular orientation. Minimizing the energy using this potential gives a total solvation energy of $-3085$ K. On the other hand, if Eq.~(\ref{min_eq}) is solved with the anisotropic He$_3^+$-He potential, a solvation energy of $-4059$ K is obtained. The librational zero-point energy correction to the latter can be estimated by considering small angular deviations around the frozen rotor solution, i.e. expanding
\begin{align}
&\varepsilon_{He^+_3-He}[\theta_{ion}] = \nonumber \\
&\int |\Psi_0(\vec{r})|^2 V_{He^+_3-He}(\vec{r}-\vec{r}_{ion},\theta_{ion})d^3r \simeq \nonumber \\
&\varepsilon_0 + \frac{1}{2}I \omega^2_\theta \theta_{ion}^2
\label{ener_expansion}
\end{align}
where $\Psi_0(r)$ is the solution of Eq.~(\ref{min_eq}). The obtained correction to the energy is $\hbar\omega_{\theta}/2 = 105$ K, which yields a total solvation energy of $-3953$ K for the non-rotating ion. The energy difference between these two cases is approximately 900 K, predicting that the non-rotating ion form prevails in the bulk. The calculated liquid distributions around the freely rotating and non-rotating ions are shown in Fig.~\ref{free-fix}. Despite the apparent geometrical differences, both configurations present very similar traits. For example, both the number of atoms in the first solvent shell and the hydrodynamic mass differ only by less than one atom. As  discussed in the next section, the energetics for helium droplets also follow a similar trend. While the intermediate hindered rotor situation is not considered here explicitly, the induced perturbation to the liquid structure must lie between the two extreme cases discussed above and, consequently, no significant changes in the relevant experimental observables are expected.

\subsection{Bosonic density functional theory calculations}

The liquid minimum energy configuration obtained from OT-DFT show a tightly bound Atkins' snowball structure around the ion, followed by several less inhomogeneous solvent layers that are practically disconnected from the snowball itself (i.e., low liquid density between them). The snowball structure contains 19 He atoms in addition to the ion itself confined inside \textit{ca.} $4.5$~\AA{} radius. This is then surrounded by the second more slowly varying solvation layer, which contains 43 additional helium atoms within a $7.7$~\AA{} radius. From this point on, the density displays much smaller variations from the bulk value and the progression finally converges towards the homogeneous bulk limit. By integrating the density in a sufficiently large simulation box and subtracting out the bulk contribution, we find that the positive charge attracts an additional 50 He atoms (including the ion). Note that a similar number was obtained by using the original Atkins'  model but here the microscopic picture is quite different. Instead of just a compact solid sphere containing the 50 atoms, the OT-DFT model shows a layered structure with the ion embedded inside a compact layer of 19 additional atoms followed by the secondary solvation shells with a total of 28 atoms more than in the homogeneous bulk.

\begin{figure}
\centering
\includegraphics[width=\linewidth]{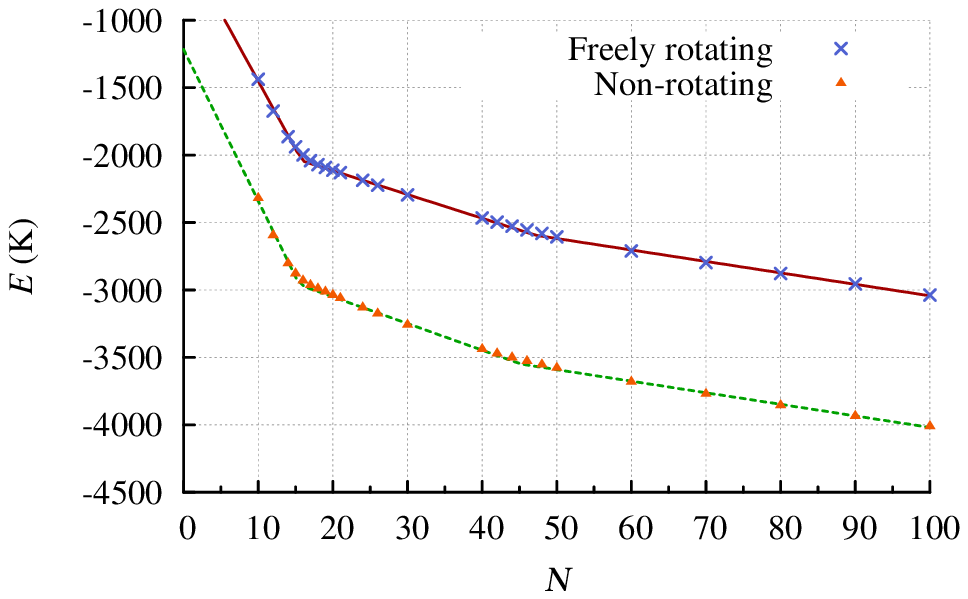}
\includegraphics[width=\linewidth]{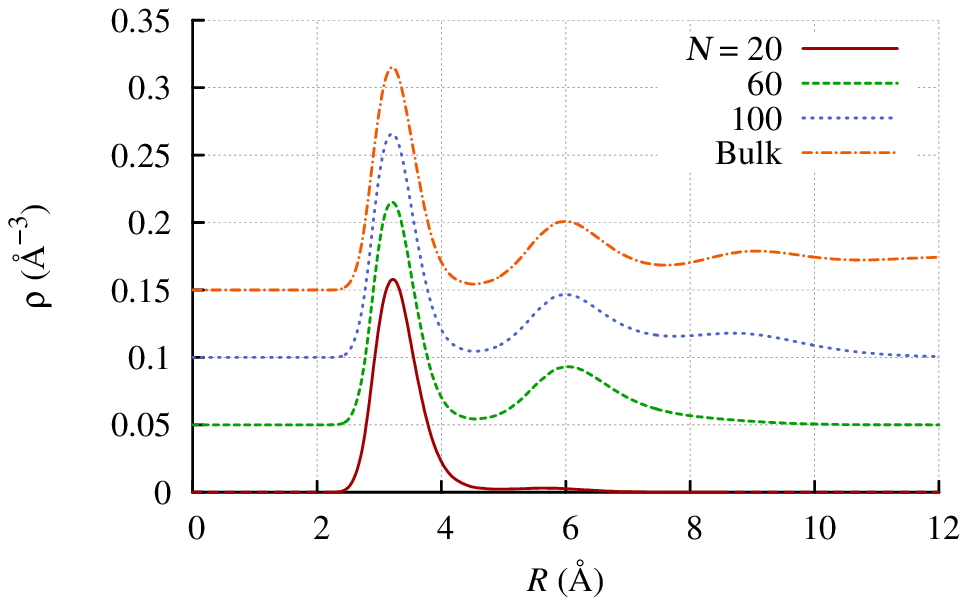}
\caption{
Top panel: Total energy of He$^+_3$+He$_N$ droplet as a function of the number of atoms ($N$) for the two rotational configurations. Bottom panel: Radial liquid densities of selected He$^+_3$+He$_N$ droplets where each density curve has been shifted by a constant on the vertical axis for the sake of clarity.}
\label{droplets}
\end{figure}

Inspection of the total energy of finite He$^+_3$+He$_N$ clusters as a function of $N$ provides information about the gradual buildup of the solvent layers. These energies are presented in the top panel of Fig.~\ref{droplets} for both freely rotating and non-rotating ions along with the radial densities for selected values of $N$ shown in the bottom panel. The dependency of the energy on the number of atoms is qualitatively the same in both cases. Even if the spatial distribution of the liquid in each configuration is different, most experimental observables are not sensitive to such atomic scale detail. Previous experiments analyzing the ionized helium droplet size distribution indicate that some values of $N$ are preferred over the others (``magic numbers'').\cite{shen1,janda1} While the results vary somewhat from experiment to experiment, the persistent magic numbers seem to appear at $N=10,14$\cite{shen1,janda1,scheier1} and several others in the range of $N=20-50$.\cite{shen1,janda1} If the appearance of such magic numbers is taken as a signature for solvent shell completion around the ion, the present OT-DFT calculations can provide important clues about their origin. The energies shown in Fig.~\ref{droplets} can be approximated by three linear regions as indicated in the figure. By correlating these regions with the density profile data shown in the bottom panel of Fig. \ref{droplets}, it can be seen that the first region ending at \textit{ca.} 16 He atoms corresponds to the completion of the first solvent shell, the second shell completes near 45 atoms and after this the system slowly evolves towards the bulk behavior. Based on this correlation, we assign the first two magic numbers at 10 and 14 to the completion process of the first solvent shell and the 20-50 atom range to the completion of the second shell. Note, however, that the most recent experiments did not show any magic numbers between 20 and 50 He atoms and therefore it is not clear if the completion of the second solvent shell has been observed experimentally.\cite{scheier1}

\subsection{The effective mass of He$_3^+$}

\begin{figure}
\includegraphics[width=\linewidth]{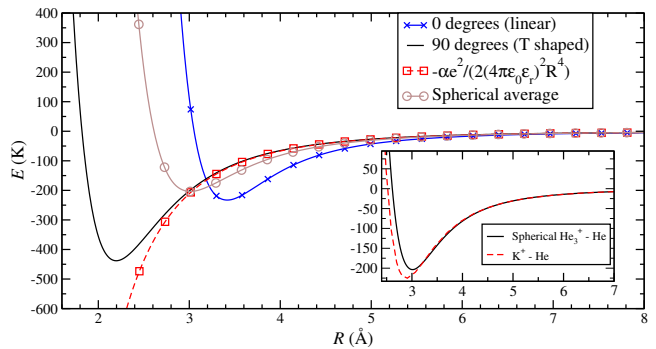}
\caption{Overview of the He$^+_3$ -- He pair potentials (linear, T-shape and spherically averaged). The charge - charge induced dipole interaction potential is also shown for reference. Inset: comparison of the spherically averaged He$_3^+$ -- He potential with the K$^+$ -- He potential used in Ref.~\citenum{eloranta2}.}\label{pot-comp}
\end{figure}

The localized liquid structure around the ion opens the question on how many atoms would follow the ion as it travels through the medium or, in other words, what is its effective mass? Even though a total of 50 atoms are attracted to the positive charge, there is a clearly defined core of 19 closely-packed atoms around He$^+_3$ that are approximately disconnected from the rest of the liquid. Our OT-DFT calculation using Eq. (\ref{frame_e}) for the non-rotating He$_3^+$ results in $m_{add} = 18.3$ $m_{He}$, which translates into an approximate $R_b$ value of 3.35~\AA{}. Both of these values are considerably smaller than the previous experimental estimates\cite{sanders1,williams2} where the effective mass was given in the range of 30 - 40 helium atoms. This rather large difference is indicative of either some unknown additional source contributing to the experimentally determined effective mass or a possible problem with the simplified models used to analyze the experimental data. Regarding the latter, while both microwave based techniques are termed ``direct measurements'', they make assumptions either about the interaction of the positive ion with liquid helium or employ phenomenological models in extracting the mass. It is possible that such models have difficulties in dealing with the chemically bound He$_3^+$ molecule interacting with the surrounding strongly inhomogeneous liquid. To quantify this statement, Fig.~\ref{pot-comp} shows a comparison between the actual He$_3^+$ -- He potentials and the simple charge - charge induced dipole interaction model. Since the positive charge forms a He$_3^+$ molecule in the liquid rather than remains as a separated charge, significant differences between the two potentials are observed. Based on this observation, the analysis behind both microwave -based techniques are better suited for the solvated electron, which remains as a well separated charge and has a very smooth liquid structure around it\cite{eloranta5,barranco1}. To demonstrate this, we have also computed the added mass for the solvated electron at saturated vapor pressure and 0 K. The OT-DFT calculation gives $m_{add} = 246$ $m_{He}$, which is in excellent agreement with the microwave based experimental data (\textit{ca.} 240 He atoms).\cite{sanders1,williams2} For the positive ion, the added mass from our OT-DFT calculation is consistent with nearly all of the first solvent shell following the ion, i.e. 18 atoms following vs. 19 atoms localized in the first solvent shell. As shown in the inset of Fig.~\ref{pot-comp}, the pair potentials between He$_3^+$(spherical) -- He and K$^+$ -- He are comparable and therefore we expect these two ions to have almost identical added masses. Our previous OT-DFT calculation for K$^+$ gave $m_{add} = 17$,\cite{eloranta2} which is in exact agreement with an independently performed quantum Monte Carlo calculation.\cite{reatto1} Furthermore, these two calculations produced nearly identical liquid density profiles for K$^+$, demonstrating that OT-DFT can capture the correlation effects even in this highly attractive system. Since both K$^+$ and He$^+_3$ have similar interaction potentials with liquid helium, they must form similar snowball structures in the liquid. The snowball radius is in turn related to the effective thermal phonon/roton scattering cross-section, which determines the ion mobility when $T>1$ K. Based on this both ions should have nearly identical mobilities, which is confirmed by the experiments of Glaberson and Johnson.\cite{johnson1} These experiments show that the ratio between He$^+$ and K$^+$ mobilities is very close to one ($\mu(He_3^+)/\mu(K^+) = 1.039$; the closest value to one for all considered alkali ions).

\section{Conclusions}

In this work, we have identified the intrinsic positive charge in superfluid as the triatomic He$_3^+$ ion, which binds ground state helium atoms through the charge - charge induced dipole interaction in a pairwise additive manner. In both helium droplets and bulk helium, this ion forms the well-known Atkins' snowball structure. However, the microscopic picture provided by the OT-DFT calculations is quite different from the Atkin's model as multiple layers of inhomogeneous liquid surrounds the ion. Evaluation of the total energy vs. helium droplet size shows three distinct linear regions, which are assigned to the completion of first and second solvent shells. These regions correlate with the earlier experimental observations of magic helium droplet sizes for ionized helium clusters. The calculated added mass in the bulk is consistent with similar calculations and experiments on alkali atoms, but deviates significantly from the previous experimental results for the positive charge. This may be a consequence of some unknown additional factor contributing to the effective mass in the experiments or a limitation of the simplified models used in analyzing the experimental data. On the other hand, both experiments and theory agree on the added mass for the negative charge, which has a much simpler solvation structure and remains as a plain distributed point charge inside the solvation bubble. In contrast, the positive ion forms a chemically bound ionic core (He$_3^+$) and the associated strongly inhomogeneous liquid density distribution.

\begin{acknowledgement}
Financial support from the National Science Foundation grants CHE-1262306 and DMR-1205734 and the Interdisciplinary Research Institute for the Sciences (IRIS) are gratefully acknowledged.

\end{acknowledgement}

\section{Appendix}

Raw data points for the ground electronic state He$_3^+$ -- He interaction as a function of $(r,\theta)$ are shown in Table \ref{tableA}. The variable $r$ is the distance between the ground state He atom and the center of mass of He$^+_3$ and $\theta$ is the angle between the linear He$^+_3$ and the approaching He. When $\theta = 0^\circ$, the system is linear and for $\theta = 90^\circ$ it is T shaped. The full potential energy surface was obtained through a fifth order polynomial interpolation for $\theta$ and spline interpolation along $r$.

\bibliography{he3+}

\onecolumn
\begin{table}
\begin{tabular}{|c|c|c|c|l|}
\cline{1-5}
He$^+_n$ & $R_1$ & $R_2$ & Symmetry & Harmonic frequencies\\
\cline{1-5}
2 (L) & 1.080 (1.081) & -- & $D_{2h}$ & $A_g$: 1698 (1698.5)\\
\cline{1-5}
3 (L) & 1.234 (1.234) & -- & $D_{2h}$ & $B_{2u}/B_{3u}$: 240 (239), $B_{1u}$: 503 (481),\\
      &               &    &          &  $A_g$: 908 (904)\\
\cline{1-5}
4 (T) & 1.234 (1.233) & 2.198 (2.208) & $C_{2v}$ & $B_2$: 92, $A_1$: 152, $A_1$: 230, $B_2$: 242,\\
      &       &       &          & $B_2$: 493, $A_1$: 907\\
\cline{1-5}
5 (X) & 1.234 (1.232) & 2.199 (2.222) & $D_{2h}$ & $B_{3u}$: 68, $B_{1g}$: 109, $A_g$: 130, $B_{2u}$: 165,\\
      &       &       &          & $B_{2u}$: 220, $B_{1u}$: 242, $B_{3u}$: 486, $A_g$: 907\\
\cline{1-5}
\end{tabular}
\caption{Overview of the calculated He$_n^+$ cluster geometries, applied molecular point groups (symmetry) and harmonic frequencies (cm$^{-1}$) with the corresponding irreducible representations. Letters L, T and X refer to the geometry of the cluster, i.e., linear, T shaped or cross shaped. The bond lengths $R_1$ and $R_2$ (\AA{}) correspond to the He$_3^+$ core and the loosely bound He atom coordinates, respectively. For He$_2^+$ $R_1$ simply corresponds to the diatomic bond length. Reference values shown in parentheses for $R_1$ and $R_2$ were taken from earlier calculations (CCSD(T)/AVQZ for He$_2^+$ and MP4(SDQ)/AVTZ for the other ions)\cite{gianturco1,yurtsever1} and for the harmonic frequencies from experiments (He$_2^+$) or from earlier QCSID(T)/AVQZ level calculations (He$_3^+$).\cite{herzberg1,gellene2}}
\label{table1}
\end{table}

\begin{table}
\begin{tabular}{|c|c|c|c|}
\cline{1-4}
$n$ / $T$ & $\Delta_r E$ & $\Delta_r H^\circ$ & $\Delta_r S^\circ$\\
\cline{1-4}
2 / 110 K & $-16.0$ ($-15.2$) & $-13.1$ ($-15.6\pm 0.6$) & $-83.6$ ($-74.1\pm 8$)\\
\cline{1-4}
3 / 40 K  & $-2.3$ ($-3.5$) & -$4.4$ ($-2.7\pm 0.6$) & $-53.9$ ($-53.8\pm 8$)\\
\cline{1-4}
4 / 30 K  & $-2.3$ ($-3.5$) & $-4.6$ ($-2.0\pm 0.6$) & $> -80$ ($-45.2 \pm 8$)\\
\cline{1-4}
\end{tabular}
\caption{Comparison of the calculated thermodynamic data for He$^+_n$ + He $\rightleftharpoons$ He$^+_{n+1}$ reaction with the experimental values. The reaction energies ($\Delta_r E$) and enthalpies are given in kJ/mol and entropies in J/(mol K) with the corresponding experimental values indicated in parentheses.\cite{mori1} Note that the calculated $\Delta_r E$ values include only the zero-point corrections to energy (i.e., $T=0$ K). The previously calculated values for $\Delta_r E$ by Scifoni and Gianturco are given as a reference but do not include the zero-point corrections for $n=3,4$.\cite{gianturco2,yurtsever1} The greater than sign signifies that the value is only a lower limit estimate (see text).}
\label{table2}
\end{table}

\begin{table}
\begin{tabular}{|c|c|c|c|c|c|}
\cline{1-6}
$r$  & $E(r,0^\circ)$        & $E(r,22.5^\circ)$     & $E(r,45^\circ)$       & $E(r,67.5^\circ)$     & $E(r,90^\circ)$\\
\cline{1-6}
2.5  &       --              & 0.5705915             & 0.1200517             & 0.0406008             & 0.0194044\\
2.6  &       --              & 0.5093558             & 0.1073081             & 0.0336180             & 0.0152284\\
2.7  &    4.5977223          & 0.4456016             & 0.0948443             & 0.0275025             & 0.0117324\\
2.8  &    2.9157765          & 0.3830794             & 0.0829540             & 0.0222467             & 0.0088439\\
2.9  &    1.9871021          & 0.3244716             & 0.0718426             & 0.0177945             & 0.0064869\\
3.0  &    1.4201165          & 0.2714513             & 0.0616352             & 0.0140662             & 0.0045863\\
3.5  &    0.3288355          & 0.1007900             & 0.0253718             & 0.0033222             & $-$0.0002899\\
3.7  &    0.1876563          & 0.0676241             & 0.0167979             & 0.0013952             & $-$0.0009637\\
3.9  &    0.1087496          & 0.0459052             & 0.0106730             & 0.0002174             & $-$0.0012809\\
4.0  &    0.0836705          & 0.0378646             & 0.0083489             & $-$0.0001729          & $-$0.0013523\\
4.1  &    0.0649785          & 0.0311506             & 0.0064299             &       --              & $-$0.0013836\\
4.2  &    0.0509560          & 0.0254952             & 0.0048574             & $-$0.0006767          & $-$0.0013847\\
4.3  &    0.0403013          & 0.0207239             & 0.0035792             & $-$0.0008261          & $-$0.0013638\\
4.4  &    0.0320563          & 0.0167094             & 0.0025488             & $-$0.0009260          & $-$0.0013271\\
4.5  &    0.0255464          & 0.0133381             & 0.0017256             & $-$0.0009873          & $-$0.0012795\\
4.7  &    0.0160948          &        --             &     --                & $-$0.0010274          & $-$0.0011660\\
4.9  &      --               &        --             &     --                & $-$0.0009982          & --\\
5.0  &    0.0075322          & 0.0033795             & $-$0.0003506          & $-$0.0009684          & $-$0.0009836\\
5.5  &    0.0012469          & 7.48$\times 10^{-5}$  & $-$0.0007716          & $-$0.0007645          & $-$0.0007153\\
6.0  & $-$0.0004883          & $-$0.0006912          & $-$0.0007034          & $-$0.0005675          & $-$0.0005169\\
6.5  & $-$0.0007344          & $-$0.0006951          & $-$0.0005435          & $-$0.0004167          & $-$0.0003781\\
7.0  & $-$0.0006120          & $-$0.0005414          & $-$0.0004017          & $-$0.0003088          & $-$0.0002816\\
7.5  & $-$0.0004513          & $-$0.0003957          & $-$0.0002961          & $-$0.0002325          & $-$0.0002138\\
8.0  & $-$0.0003248          & $-$0.0002874          & $-$0.0002211          & $-$0.0001782          & $-$0.0001652\\
9.0  & $-$0.0001750          & $-$0.0001593          & $-$0.0001301          & $-$0.0001098          & $-$0.0001033\\
10.0 & $-$0.0001028          & $-9.57\times 10^{-5}$ & $-8.18\times 10^{-5}$ & $-7.14\times 10^{-5}$ & $-6.79\times 10^{-5}$\\
11.0 & $-6.49\times 10^{-5}$ & $-6.14\times 10^{-5}$ & $-5.41\times 10^{-5}$ & $-4.84\times 10^{-5}$ & $-4.65\times 10^{-5}$\\
12.0 & $-4.33\times 10^{-5}$ & $-4.14\times 10^{-5}$ & $-3.74\times 10^{-5}$ & $-3.40\times 10^{-5}$ & $-3.29\times 10^{-5}$\\
13.0 & $-3.02\times 10^{-5}$ & $-2.90\times 10^{-5}$ & $-2.66\times 10^{-5}$ & $-2.46\times 10^{-5}$ & $-2.39\times 10^{-5}$\\
14.0 & $-2.17\times 10^{-5}$ & $-2.10\times 10^{-5}$ & $-1.95\times 10^{-5}$ & $-1.83\times 10^{-5}$ & $-1.78\times 10^{-5}$\\
15.0 & $-1.60\times 10^{-5}$ & $-1.56\times 10^{-5}$ & $-1.47\times 10^{-5}$ & $-1.38\times 10^{-5}$ & $-1.35\times 10^{-5}$\\
16.0 & $-1.21\times 10^{-5}$ & $-1.19\times 10^{-5}$ & $-1.12\times 10^{-5}$ & $-1.07\times 10^{-5}$ & $-1.05\times 10^{-5}$\\
17.0 & $-9.4\times 10^{-6}$  & $-9.2\times 10^{-6}$  & $-8.7\times 10^{-6}$  & $-8.4\times 10^{-6}$  & $-8.2\times 10^{-6}$\\
18.0 & $-7.4\times 10^{-6}$  & $-7.2\times 10^{-6}$  & $-6.9\times 10^{-6}$  & $-6.6\times 10^{-6}$  & $-6.5\times 10^{-6}$\\
19.0 & $-5.9\times 10^{-6}$  & $-5.8\times 10^{-6}$  & $-5.5\times 10^{-6}$  & $-5.3\times 10^{-6}$  & $-5.3\times 10^{-6}$\\
20.0 & $-4.7\times 10^{-6}$  & $-4.7\times 10^{-6}$  & $-4.5\times 10^{-6}$  & $-4.3\times 10^{-6}$  & $-4.3\times 10^{-6}$\\
25.0 & $-1.9\times 10^{-6}$  & $-1.9\times 10^{-6}$  & $-1.8\times 10^{-6}$  & $-1.8\times 10^{-6}$  & $-1.8\times 10^{-6}$\\
30.0 & $-9\times 10^{-7}$    & $-9\times 10^{-7}$    & $-9\times 10^{-7}$    & $-9\times 10^{-7}$    & $-9\times 10^{-7}$\\
50.0 & $-1\times 10^{-7}$    & $-1\times 10 ^{-7}$   & $-1\times 10^{-7}$    & $-1\times 10^{-7}$    & $-1\times 10^{-7}$\\
\cline{1-6}
\end{tabular}
\caption{Raw data points obtained from RCCSD(T)/AV6Z calculations as a function of $(r,\theta)$. The distance $r$ is expressed in Bohr, $\theta$ in degrees and $E(r,\theta)$ in Hartree. Entries marked with '--' were not computed as they were not needed for the interpolation procedure.}
\label{tableA}
\end{table}

\end{document}